\documentclass[superscriptaddress,prl, floatfix, reprint,amsmath,amssymb, aps]{revtex4-1}
\usepackage{graphicx}
\usepackage[utf8]{inputenc}
\usepackage[T1]{fontenc}

\newcommand{\tp}{{t^\prime}}
\newcommand{\tpp}{{t^{\prime\prime}}}

\newcommand{\xp}{{x^\prime}}

\newcommand{\yp}{{y^\prime}}

\newcommand{\ypp}{{y^{\prime\prime}}}

\newcommand{\eq}{\mathrm {eq}}
\newcommand{\xav}{x_{\mathrm {ave}}}
\newcommand{\yav}{y_{\mathrm {ave}}}

\newcommand{\erfc}{\mathrm{erfc}}

\begin{document}

\title{
 Fredholm theory for the mean first-passage time of integrate-and-fire oscillators with colored noise input
}

\author{Carl van Vreeswijk} 
\affiliation{Centre de Neurophysique Physiologie et
Pathologie, Paris Descartes University and CNRS UMR 8002 INCC, 75006 Paris, 
France}

\author{Farzad Farkhooi} 
\email{Corresponding author: farzad@bccn-berlin.de}
\affiliation{Institute for Theoretical Biology, Department of Biology,
Humboldt-Universit\"at zu Berlin, 10115 Berlin, Germany}

\date{\today}

\begin{abstract}
We develop a method to investigate the effect of noise timescales on the
first-passage time of nonlinear oscillators. Using Fredholm theory, we derive
an exact integral equation for the mean event rate of a
leaky-integrate-and-fire oscillator that receives constant input and temporally
correlated noise.  Furthermore, we show that Fredholm theory provides a unified
framework to determine system scaling behavior for small and large noise
timescales. In this framework, the leading order and higher-order asymptotic
corrections for slow and fast noise are naturally emerging. We show the scaling
behavior in the both limits are not reciprocal.  We discuss further how this
approach can be extended to study the first-passage time in a general class of
nonlinear oscillators driven by colored noise at arbitrary timescales.
\end{abstract}

\maketitle

The dynamics of nonlinear oscillators that receive temporally correlated inputs
plays a central role in the analysis of many physical, chemical and biological
systems \cite{wax_selected_1954, *kampen_stochastic_2007,
*risken_fokker-planck_1996}.  The standard method that is used to treat
stochastic dynamics that are governed by temporally correlated noise is to
approximate the probability law of the system using a Fokker-Planck-like
evolution equation (FPE) \cite{mcClintock_book_1989, haunggi_colored_1994,
doering_bistability_1987, *klosek_colored_1998, *hagan_explicit_1999}.  Most
existing analysis is only applicable when the noisy input correlation time is
either much shorter or much longer than the oscillator intrinsic timescale
\cite{haunggi_colored_1994}.  Many challenging and interdisciplinary questions
remain regarding the analysis of stochastic dynamics in the case of nonlinear
systems with noise that has intermediate timescales.

In this letter, we determine the mean event rate of a specific nonlinear
oscillator: a leaky-integrate-and-fire (LIF) neuron that receives input which
fluctuates over arbitrary timescales. Although the LIF is used widely in the
mathematical description and numerical simulations of neural circuits
\cite{brunel_lapicques_2007, *teeter_generalized_2018}, a precise analytical
approach that yields the exact system response at all correlation timescales
has not yet been developed. We use a mathematical approach that is based on
Fredholm theory \cite{fredholm_sur_1903} to address this gap. Our method yields
an effective transfer function in the form of an integral equation for
arbitrary noise correlation times. Furthermore, our method readily provides an
asymptotic expansion term for system limiting behavior in the fast noise case,
similar to previous results \cite{brunel_firing_1998, *fourcaud_dynamics_2002,
*schuecker_modulated_2015}. We also, for the first time, derive an asymptotic
expansion in the slow noise limit. Our results enable us to understand the
interplay between finite noise timescales when shaping nonlinear system
dynamics.

We consider the dynamics of an LIF oscillator whose membrane voltage, $x$, and
input variable, $y$, satisfyies
\begin{eqnarray} \frac{d}{dt} x & = & \alpha_m \left[\mu -x +
\sqrt{\frac{\alpha_m+\alpha_s}{\alpha_m}}\sigma y\right] \nonumber \\
\frac{d}{dt} y & = & -\alpha_s y+ \sqrt{\alpha_s} \eta(t), \label{system}
\end{eqnarray} where $\alpha_m=1/\tau_m$ and $\alpha_s=1/\tau_s$. Here,
$\tau_m$ is the membrane time constant, $\tau_s$ is the noise correlation time,
$\eta(t)$ is the white noise random variable, and $\sigma$ is the noise
amplitude.  An oscillator emits an event whenever the membrane reaches the
threshold, $x(t^-)= x_{th}=1$; in this case, the voltage returns immediately to
the resting potential (reset), $x(t^+)=0$. The input scaling factor
$\sqrt{\frac{\alpha_m+\alpha_s}{\alpha_m}}$ ensures that the input fluctuation
does not die out in the limit $\tau_s\rightarrow\infty$, and the equilibrium
distribution of $y$ is $P_y= e^{-y^2}/\sqrt{\pi}$. Additionally, in the absence
of an event as $x_{th} \to \infty$, the equilibrium distribution of $x$ is
independent of $\alpha_s$ and is given by $P_{x} =
e^{-(x-\mu)^2/\sigma^2}/(\sqrt{\pi}\sigma)$\@.

The standard approach to analyzing Eq. (\ref{system}) is to study its FPE, as
follows: 
\begin{eqnarray} \frac{\partial}{\partial t} \rho(x,y,t)  =
&-&\frac{\partial}{\partial x}J_x(x,y,t) -\frac{\partial}{\partial y}J_y(x,y,t)
\nonumber \\ &+&\big[\delta(x)-\delta(x-1)\big] r(y,t) , \label{FKP}
\end{eqnarray}
where $\rho(x,y,t)$ is the probability density of the system being in the state
$(x,y)$ at time $t$, $J_x(x,y,t)= \aleph_x \, \rho(x,y,t)$ and
$J_y(x,y,t)=\alpha_s\left[-y-\frac{1}{2}\frac{\partial}{\partial y}
\right]\rho(x,y,t)$ are the flux in $x$ and $y$, respectively, and $\aleph_x
\equiv \alpha_m\left(\mu-x+ \sqrt{\frac{\alpha_m+\alpha_s}{\alpha_m}}\sigma\,
y\right)$.
The reset rate, $r(y,t)$, is the rate at which $x$ reaches the threshold
($x=1$) at noise level $y$ and at time $t$; $r(y,t)$ is given by 
\begin{equation} r(y,t)=J_x(1,y,t).  \end{equation}
This system of equations has proven difficult to solve directly due to
complications associated with the reset mechanism.

\begin{figure}[ht!] \centering \includegraphics[scale=1.00]{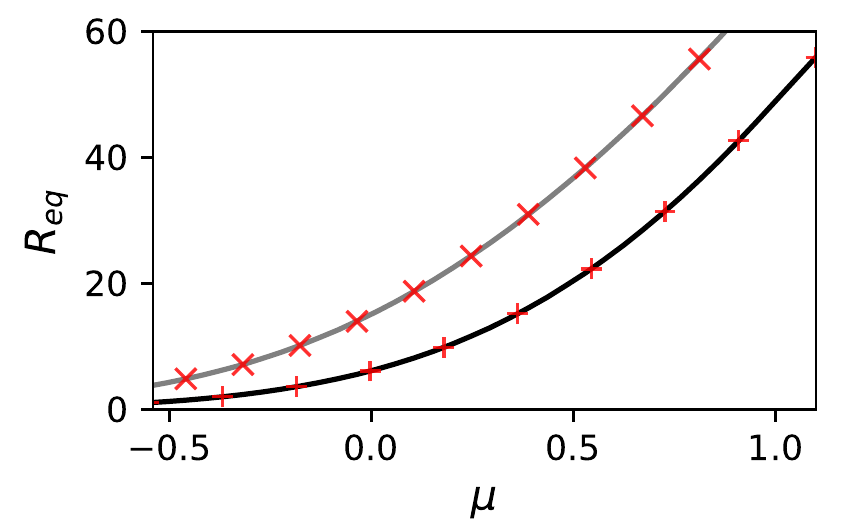}
	\caption{Event rate of an LIF oscillator as a function of mean input
	for two different synaptic filtering dynamics; the \textbf{black} line
	indicates $\alpha_s=1000$ and the \textbf{grey} line indicates
	$\alpha_s=10$. Symbols are simulation results of an LIF for 1000
	trials; error bars are smaller than symbol size. Parameters:
	$\sigma=1.0$, $\alpha_m=100.0$.} \label{gain} \end{figure}

\begin{figure}[t!] \centering \includegraphics[scale=1.]{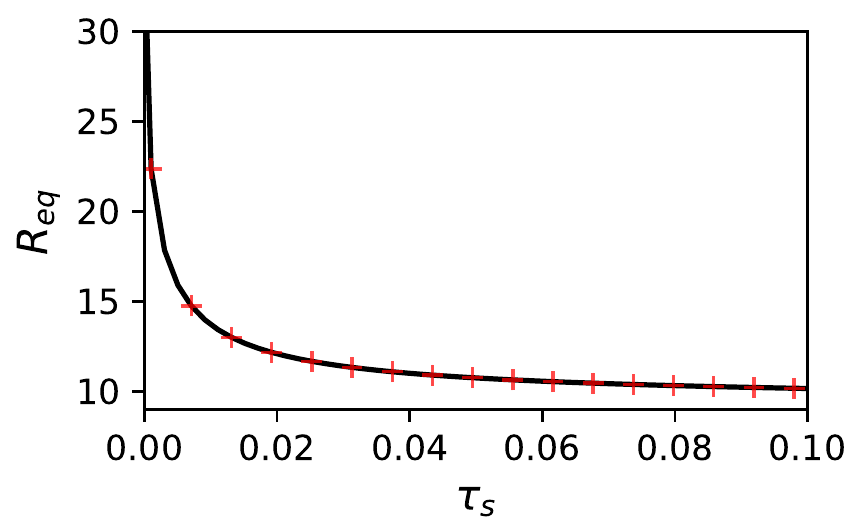}
	\caption{Event rate of an LIF oscillator as a function of synaptic
	timescale. The \textbf{black} line indicates the output rate obtained by
	solving Fredholm equation in Eq.(\ref{fredholm:eq}). The
	\textbf{crosses} indicate the event rate resulting from numerical
	simulations of an LIF oscillator over 1000 trials, as described in Eq.
	\ref{system}. Parameters: $\mu=0.2$,  $\sigma=1.0$, $\alpha_m=100.0$.
} \label{r_eq} \end{figure}

To \textit{resolve} this challenging problem,  let
$\hat{\rho}(x,y,t|\xp,\yp,\tp)$ be the probability density of state variables
$(x,y)$ at time $t$ in the absence of a spiking mechanism, given the initial
condition $(\xp, \yp)$ at time $\tp$. Using system invariance under time
translation, we observe that this unrestricted probability density is a
function of $t-\tp$, and that it can be written as
$\hat{\rho}(x,y,t-\tp|\xp,\yp)$\@. Moreover, the system without the reset
mechanism is simply a linear set of stochastic differential equations with a
Gaussian noise variable (Eq. (\ref{system})).  Thus, $\hat{\rho}$ is completely
determined by its mean  $(\xav,\yav)$ and its covariance matrix ($C$); details
are given in Ref.~\cite{supp}.  To include the reset mechanism, we must remove
the oscillator  at $x=1$ and re-insert it at $x=0$, keeping the value of $y$
unchanged at time $\tpp$ with a rate of  $r(y,\tpp-\tp|\xp,\yp)$.  Therefore,
the probability density, $\rho(x,y,t|\xp,\yp,\tp)$  (Eq.~(\ref{FKP})) for the
oscillator state to be $(x,y)$ at time $t$, given that the state was
$(\xp,\yp)$ at time $\tp$, is given by 
\begin{eqnarray} \rho(x,y,t&|&\xp,\yp,\tp)  = \hat{\rho}(x,y,t-\tp|\xp,\yp)
	\nonumber \\ &-&\int_\tp^t\!d\tpp\! \int\! d\ypp \,
	[\hat{\rho}(x,y,t-\tpp|1,\ypp) \nonumber \\ &- &
	\hat{\rho}(x,y,t-\tpp|0,\ypp)] r(\ypp,\tpp-\tp|\xp,\yp).
	\label{rho:eq} \end{eqnarray}
Note that Eq.\eqref{rho:eq} is exact; because Eq.\eqref{FKP} is a linear PDE with the boundary conditions inhomogeneity and Eq.\eqref{rho:eq} is its solution based on its Green's function (propagator) \cite{risken_fokker-planck_1996}. 
The rate of oscillator removal at the threshold is, indeed, where $x$ fluxes
through $x=1$ from below the threshold. Thus,
\begin{align} r(y,t-\tp|\xp,\yp)&=\left[\aleph_1 \, \rho(1,y,t-\tp|\xp,\yp)
\right]_+, \label{r:eq} \end{align}
where $[.]_+$ is a half-rectification function. Note that, since $\rho$ is
non-negative, we obtain $r=0$ for $y<y_- \equiv
\sqrt{\frac{\alpha_m}{\alpha_m+\alpha_s}}z(1)$, where $z(x)\equiv
\frac{(x-\mu)}{\sigma}$.
Taking Eq. (\ref{r:eq}) and inserting $\rho(x,y,t-\tp)$ for  $x=1$ from Eq.
(\ref{rho:eq}) yields a self-consistency equation for $r(y,t-\tp|\xp,\yp)$. 
We take the limit $\tp\rightarrow-\infty$ to obtain the equilibrium value for
$r$;  $r(y,t-\tp|\xp,\yp)$ and $\hat{\rho}(x,y,t-\tp|\xp,\yp)$ both reach
steady state values in this limit. Note that $r_\eq(y)$ and
$\hat{\rho}_\eq(x,y)$ are independent of $\xp$ and $\yp$, respectively.
Furthermore, since $\hat{\rho}(x,y,t|1,\yp)-\hat{\rho}(x,y,t|0,\yp)$ decays as
$e^{-\alpha_m t}$ for large $t$, we obtain 
\begin{eqnarray} r_\eq(y)&=& \aleph_1
\left(\hat{\rho}_\eq(1,y)-\int_{y_-}^\infty\!d\yp\,K(y,\yp)r_\eq(\yp) \right),
\label{fredholm:eq} \end{eqnarray} where the kernel, $K$, is given by
\begin{equation} K(y,\yp)=\int_0^\infty\!\! dt\,\,
\left[\hat{\rho}(1,y,t|1,\yp)-\hat{\rho}(1,y,t|0,\yp) \right].
\label{kernel}\end{equation}
Eq. (\ref{fredholm:eq}) is the Fredholm equation of the second kind. Since we
already have expressions for both $\hat{\rho}_\eq$ and the kernel $K$ (details
are given in Ref.~\cite{supp}), Eq. (~\ref{fredholm:eq}) uniquely determines
$r_\eq(y)$  for $y\geq y_-$. Finally, the output event rate, $R_\eq$, which
describes that rate at which an oscillator emits spikes at equilibrium, is
given by 
\begin{equation} R_\eq=\int_{y_-}^\infty\! dy\, r_\eq(y).  \end{equation}
The solution to this equation is easily obtained numerically using standard
techniques (details are given in Ref.~\cite{supp}).  Figs.\ref{gain} and
\ref{r_eq} illustrate the dependence of the mean event rate on various
parameters. In Fig. \ref{gain}, the output rate, $R_\eq$, is plotted against
mean input, $\mu$, for two intermediate values of $\tau_s$; the rate is reduced
for larger $\tau_s$. Fig. \ref{r_eq} demonstrates the dependence of $R_\eq$ on
arbitrary $\tau_s$. We observe that the event rate is strongly dependent on
synaptic filtering. Fredholm theory for the escape rate (Eq.
(\ref{fredholm:eq})), presented here, also allows analytical study of the
asymptotic behavior, in both the fast and slow noise regimes.

\begin{figure}[t!] \centering \includegraphics[scale=1.]{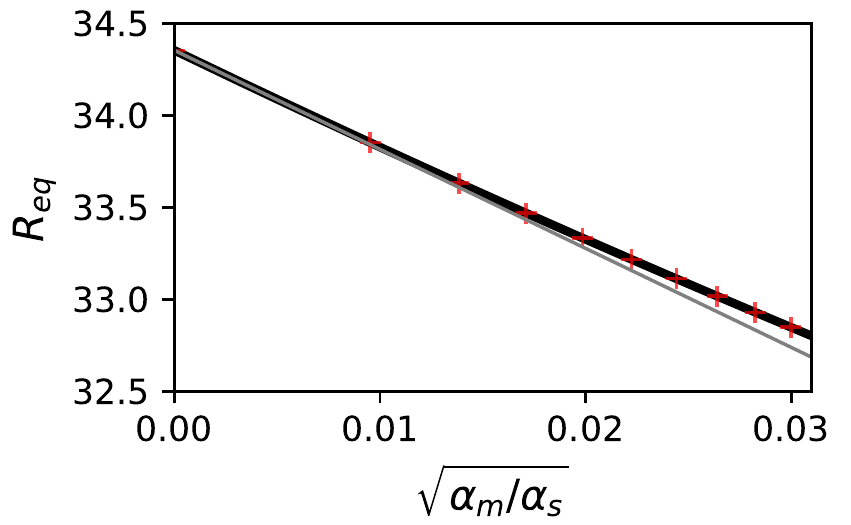} \caption{
		Fast noise regime for the event rate of an LIF oscillator as a
		function of $\sqrt{\frac{\alpha_m}{\alpha_s}}$. The \textbf{black
		thick} line displays the full solution of
		Eq.(\ref{fredholm:eq}) and the \textbf{thin grey} line
		displays the linear approximation of firing for large and
		finite $\alpha_s$. \textbf{Crosses} are simulations of an LIF
		for $10^5$ trials with a duration of $1000\times\tau_m$.
		Parameters: $\mu=.20$,  $\sigma=1.$, $\alpha_m=100.0$. }
	\label{fig3} \end{figure}

To determine the asymptotic correction for \textit{the fast noise regime}, we
must expand $\hat{\rho}_\eq(1,y)= \sum_{n=1}^{\infty}
\left(\frac{\alpha_m}{\alpha_s}\right)^{n/2} \hat{\rho}_{f_n}^{\eq}(1,y)$ and $
K(y,\yp)=\sum_{n=1}^{\infty} \left(\frac{\alpha_m}{\alpha_s}\right)^{n/2}
K_{f_n}(y,\yp)$. We make the Ansatz that $r_\eq(y)=\sum_{n=0}^{\infty}
(\frac{\alpha_m}{\alpha_s})^{n/2} r_{f_n}(y)$. We obtain 
\begin{equation} \sum_{n=0}^{\infty}
\left(\frac{\alpha_m}{\alpha_s}\right)^{n/2} r_{f_n}(y) = \alpha_m
\sum_{n=-1}^{\infty} \left(\frac{\alpha_m}{\alpha_s}\right)^{n/2} F_n(y)
\label{FH_fast1} \end{equation}
where, $F_n(y)$ collects terms of order $(\frac{\alpha_m}{\alpha_s})^{n/2}$.
Since the right-hand side of Eq. (\ref{FH_fast1}) only has terms with $n \geq
0$, we must to impose that $F_{-1}(y)=0$ for $y \geq y_-$. Therefore, as shown
in Ref.~\cite{supp}, to leading order, the event rate, $R_{f_0}\equiv
\int_{y_-}^\infty dy \, r_{f_0}(y)$, is given by $R_{f_0} =
\frac{\alpha_m}{I_R(z(0), z(1))},$ where $ I_R(z_0, z_1) = 2 \int_{z_1}^{z_0}
dz e^{z^2} \int_z^{\infty} dz' e^{-(z')^2}$.  This is, indeed, the firing rate
of an LIF neuron receiving white noise input \cite{ricciardi_diffusion_1977}.
To obtain the first order asymptotic correction to the white noise case, we
must evaluate $F_0(y)$ in Eq. (\ref{FH_fast1}); this gives the Fredholm theory
for the first order correction. Using the linearity of the Fredholm operator
and its resolvent properties in Eq. (\ref{FH_fast1}) for $n=1$ (details are
given in Ref.~\cite{supp}), we can write the asymptotic correction of the fast
noise limit as
\begin{equation} R_{f_1} = -\frac{\alpha_m}{\Phi_0 I(z(0), z(1))^2}
J_R(z(0),z(1)), \label{R_f1} \end{equation}
 where $J_R(z_0, z_1)=
 2\sqrt{\pi}(\exp({z_0}^2)\,\erfc(z_0)-\exp({z_1}^2)\,\erfc(z_1))$ and $\Phi_0
 = \frac{-\sqrt{2}}{\zeta(\frac 1 2)}$ (up to $10^{-10}$ numerical accuracy,
 see Ref.~\cite{supp} for details),  where $\zeta$ is the Riemann zeta
 function.
This is consistent with previous results \cite{brunel_firing_1998,
fourcaud_dynamics_2002,schuecker_modulated_2015}, that use boundary layer and
half-range expansion theories \cite{doering_bistability_1987,
klosek_colored_1998, hagan_explicit_1999}. Interestingly, the constant $\Phi_0$
corresponds to Milne extrapolation lengths for the FPE \cite{r_unsolved_1997}.
The Eq.\eqref{R_f1} yields the linear rate correction  $R_\eq = R_{f_0} +
\sqrt{\frac{\alpha_s}{\alpha_m}} R_{f_1} $ in the fast noise limit.
Fig.~\ref{fig3} demonstrates the limiting behavior of the event rate in the
near white noise regime; the full solution of the Fredholm equation using
Eq.(\ref{fredholm:eq}) (tick red line) and linear asymptotic correction
according to Eq.(\ref{R_f1}) (thin grey line) are plotted against
$\sqrt{\frac{\alpha_m}{\alpha_s}}$. The simulation results shown in
Fig.~\ref{fig3} (cross symbols) provide an excellent agreement with the full
solution (thick black line).

\begin{figure}[t!] \centering \includegraphics[scale=1.]{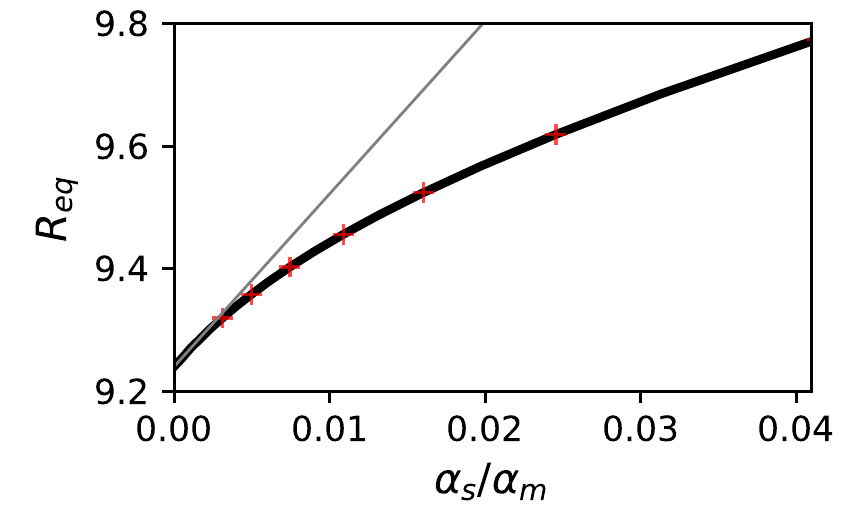} \caption{
		Slow noise regime for the event rate of an LIF oscillator as a
		function of $\frac{\alpha_s}{\alpha_m}$. The \textbf{thick black}
		line is the full solution of Eq.(\ref{fredholm:eq}) and the
		\textbf{thin grey} line is its the linear approximation for
		small and finite $\alpha_s$. Mean event rates of numerical
		simulations of an LIF oscillator for $10^5$ trials with a
		duration of $1000/\alpha_s$ are indicated by \textbf{crosses}.
		Parameters: $\mu=0.2$,  $\sigma=1.0$, $\alpha_m=100.0$. }
	\label{fig4} \end{figure}

The asymptotic correction in \textit{the slow noise regime} is also a
straightforward application of a perturbation calculation in our approach. In
the slow noise limit ($\alpha_s\rightarrow 0$), we can assume that the level of
noise is constant between two neighboring events and the inter-event-interval
is $t_0(y) \equiv\alpha_m^{-1}[\log(\mu+\sigma y)- \log(\mu+\sigma y-1)]$ for
$y>y_-$ \cite{moreno_response_2002, *moreno-bote_role_2004,
*moreno-bote_response_2010}. Therefore, to leading order, $r_{s_0}(y)$  is
given by \begin{eqnarray} r_{s_0}(y) & = & \frac{P_\eq({y})}{t_0(y)} =
\frac{\alpha_m}{\sqrt{\pi}}e^{-y^2} \left[\log\left(\frac{\mu+\sigma
y}{\mu+\sigma y-1}\right)\right]^{-1} \label{parga:eq} \end{eqnarray}
 for $y>y_-$, and $r_{s_0}(y)=0$ otherwise.  
Although this result is an already established
\cite{moreno_response_2002,*moreno-bote_role_2004, *moreno-bote_response_2010},
to the best of our knowledge, asymptotic correction terms for non-zero but
small $\alpha_s$ have not yet been determined. To simplify the calculation, we
rescale the noise to be independent of $\alpha_s$ by setting $\Sigma \equiv
\sqrt{\frac{\alpha_m+\alpha_s} {\alpha_m}}\sigma$; dependence on $\alpha_s$ can
be re-introduced at a later stage. To determine the first order correction in
the slow noise case, we observe that, for
$\frac{y-y_-}{\sqrt{\frac{\alpha_s}{\alpha_m}}}\gg 1$, $\hat{\rho}_\eq(1,y)$ is
exponentially small and can be neglected and the kernel $K(y,\yp)$ is
exponentially small unless $\yp-y$ is of order $\sqrt{\alpha_s/\alpha_m}$\@.
Therefore,  for $\frac{y-y_-}{\sqrt{\alpha_s/\alpha_m}}\gg 1$ we have $\yp\sim
y+\sqrt{\frac{\alpha_s}{\alpha_m}}Y$ in Eq. \ref{fredholm:eq}  and using the
Taylor expansion in $Y$ of $r_\eq(y+\sqrt{\alpha_s/\alpha_m}Y)$ we can rewrite
$r_\eq$  as 
\begin{equation} r_\eq(y) =\beth \left(\sum_{n=0}^\infty
\left[\frac{\alpha_s}{\alpha_m}\right]^{n/2}K_n(y) \frac{d^n}{dy^n}\right)
r_\eq(y), \label{expand1:eq} \end{equation} 
where $\beth = -\alpha_m[\mu-1+\Sigma y]$, as given in Ref.~\cite{supp}, and
$K_n$  must be expanded as $ K_n(y)=\sum_{k=0}^{\infty}
\left[\frac{\alpha_s}{\alpha_m}\right]^{k/2}K_{n,k}(y)$, where $K_{n,k}$ are
independent of $\alpha_s/\alpha_m$\@. Importantly, $K_{n,k}(y)=0$ when  $n+k$
is odd and  $\beth K_{0,0}(y)=-1$ (details are given in Ref. \cite{supp}).
Inserting this in Eq.(\ref{expand1:eq}), we obtain 
\begin{equation} \sum_{m=1}^\infty \left[
\frac{\alpha_s}{\alpha_m}\right]^{m-1} \sum_{n=0}^{2m}
K_{n,2m-n}\frac{d^n}{dy^n}  r_\eq(y) =0.  \label{expand2:eq} \end{equation}
Interestingly, because $K_{n,m}=0$ when $n+m$ is odd, the leading order
correction is of order $\alpha_s/\alpha_m$ rather than
$\sqrt{\alpha_s/\alpha_m}$\@. Thus, we expand $r_\eq$ in powers of
$\alpha_s/\alpha_m$ as 
\begin{equation} r_{eq}(y)=\sum_{n=0}^\infty \left[
\frac{\alpha_s}{\alpha_m}\right]^n r_{s_n}(y).  \label{rexp:eq} \end{equation} 
Inserting Eq. (\ref{rexp:eq}) into Eq. (\ref{expand1:eq}) and collecting terms
with the same power of $\alpha_s/\alpha_m$, we find that $r_{s_n}$ satisfies
\begin{equation} \mathcal{K}r_{s_n}(y)=-S_n(y), \label{slowlead:eq}
\end{equation} 
where the operator is given by $\mathcal{K}\equiv
K_{0,2}(y)+K_{1,1}(y)\frac{d}{dy}+ K_{2,0}\frac{d^2}{dy^2}$ , and for $n\geq1$,
$S_n$ satisfies 
\begin{equation} S_n(y)=\sum_{k=0}^{n-1}\left[\sum_{\ell=0}^{2(n+1-k)}
K_{i,2(n+1-k)-\ell}(y)\frac{d^\ell}{dy^\ell}\right]r_{s_k}(y), \end{equation} 
and $S_0(y)=0$\@.
Since $\mathcal{K}$ is a second order differential operator, Eq.
(\ref{slowlead:eq}) determines $r_{s_n}$ up to two integration constants,
provided that all $r_{s_k}$ for $k\in\{0,1,\ldots,n-1\}$ are given. This does
not completely determine $r_{s_n}$ because we have only considered $r_\eq$ for
$\frac{y-y_-}{\sqrt{\alpha_s/\alpha_m}}\gg 1$\@.  
However, we can still determine the asymptotic corrections since we can write
the scaling factor $c_n\equiv{r_{s_n}}/r_{s_0}$ and insert it into
Eq.~\ref{slowlead:eq} and thus $c_n$ satisfies 
\begin{equation}\left[\frac{d}{dy}-2y\right] \frac{d}{dy} c_n(y)=-s_n(y),
\label{scale} \end{equation}
where $s_n(y)=t_0(y)S_n(y)/(K_{2,0}(y)P_\eq(y))$.
This is clearly consistent with $c_0(y)=1$ in $r_{s_0}(y)=c_0(y)
P_\eq(y)/t_0(y)$.
For large $y$ and $\frac{y-y_-}{\sqrt{\alpha_s} t_0(y)}\gg 1$, the kernel
$K(y,y')$ becomes exponentially small; therefore, as $y\rightarrow\infty$,
fluctuations in $y$ are negligible for any order $n$. Hence, for $n>0$, $c_n(y)
\rightarrow 0$ and $\frac{d}{dy}c_n(y)\rightarrow 0$ as $y\rightarrow\infty$.
Thus, $c_n(y)$ satisfies
\begin{equation} c_n(y)= \int_y^\infty\!dy_1 e^{y_1^2}
\int_{y_1}^\infty\!dy_2\, s_n(y_2) e^{-y_2^2}.  \label{c_1} \end{equation}
This determines the leading order correction, $r_{s_1}(y)=c_1(y)r_{s_0}(y)$ and
$s_1$ is given in Ref. \cite{supp}. 
Here, we obtain Eq.\eqref{c_1} assuming that $\Sigma$ is constant, so the
scaling factor can be reformulated as
$\tilde{c}_1=c_1-\frac{y}{2}(\frac{d}{dy}t_0(y))/t_0(y)$ to return to the
original formulation of the problem.  Now, using
$R_{s_1}\equiv\int_{y_-}^{+\infty} dy' r_{s_1}(y')$ we obtain $R_{\eq}= R_{s_0}
+ \frac{\alpha_s}{\alpha_m} R_{s_1}$. Fig. \ref{fig4} illustrates the linear
approximation (thin grey line) of the event rate for small but finite
$\alpha_s$ tangents to the full solution of Fredholm equation (thick black line)
in Eq.(\ref{fredholm:eq}).

In this letter, we studied the nonlinear dynamics of an LIF oscillator that is
driven by colored noise. We derived, for the first time, an exact expression
for the event rate of the model for arbitrary correlation times in the form of
a Fredholm equation, which can readily be evaluated numerically. This approach
does not require the separation of timescales and weak noise expansion that are
typically assumed in the classical analysis of colored noise in stochastic
dynamics \cite{haunggi_colored_1994, mcClintock_book_1989}. Additionally, we
show that Fredholm theory provides a uniform formalism by which to
systematically calculate the fast and slow noise asymptotic expansions. These
expansions lead to the interesting conclusion that the system exhibits
different scaling behaviors in slow and fast noise regimes. Most previous works
in the fast noise regime use boundary-layer theory to derive the leading order
correction to the mean rate \cite{doering_bistability_1987,
klosek_colored_1998, hagan_explicit_1999, brunel_firing_1998,
fourcaud_dynamics_2002, schuecker_modulated_2015}. Our approach recovers this
result. Formally, application of FPE boundary layer theory requires the
assumption that the potential well is smooth and has zero slope at the
absorbing upper boundary. Remarkably, our result indicates that the details of
the potential do not contribute to the correction term. In the slow noise
extreme ($\alpha_s\rightarrow 0$), Moreno-Bote et al. \cite{
	moreno_response_2002, *moreno-bote_role_2004,
	*moreno-bote_response_2010} used an adiabatic approach to derive the
mean event rate; we have derived the same result. It is noteworthy, Moreno-Bote et al. \cite{moreno-bote_theory_2008} showed that in the limit of large $\tau_s$ and an additional white noise the leading order correction is linear. The unified framework here allows to generalize their results systematically and also calculate the magnitude of the slow noise correction. 
Our analysis shows that the order of the asymptotic corrections at
the both slow and fast noise timescales do not scale reciprocally; the order of
limiting behavior for the case of fast noise is $\sqrt{\tau_s/\tau_m}$, while
for slow noise it is $\tau_m/\tau_s$. Our asymptotic analysis for large
and small $\alpha_s$ indicates that linear regimes are fall outside the
physiological relevant range of synaptic dynamics (Figs. ~\ref{fig3} and
~\ref{fig4}). This demonstrates the importance of the full solution of the
Fredholm equation for the investigation of neural network dynamics.

Our approach can be extended to calculate the response of LIF
units to infinitesimal non-stationarities in the input. This can be used to
evaluate the stability of an asynchronous state of recurrent networks. To this end, one needs to  
follow the perturbation theory developed in \cite{farkhooi_renewal_2015}. Furthermore,  using Markovian embedding method one can consider a non-exponentially correlated temporal input (for small noise, $\sigma/\mu \ll 1$) \cite{haunggi_colored_1994} similar to work by Schwalger et al.\cite{schwalger_statistical_2015} for the perfect-integrate-and-fire neurons.

Our method can be applied when the solution to the unrestricted process, $\hat{\rho}$, is known. For example, our method can be used in the normative models of decision-making in a dynamic environment that an agent values recent  observations  more than older one \cite{ossmy_timescale_2013}; in the case of exponential discounting of the observations, one can directly apply our results. The other interesting example is Kubo's stochastic model that describes a irreversible process in which the noise variable takes discrete values with a Poisson switching. In Kubo's model $\hat{\rho}$ is readily determined for an arbitrary drift term \cite{kubo_stochastic_2007, haunggi_colored_1994}. This model has been used extensively in analyzing the kinetic theory of gases and the statistical theory of line-broadening \cite{saven_molecular_1993,*bezzerides_theory_1969}. In cases where oscillator dynamics can be described by a motion equation of phase variable, a Fourier expansion of $\hat{\rho}$ is typically available \cite{hongler_exact_1982}. In this case, an arbitrary-precise solution can be constructed by considering the first $n$ Fourier moments as it has been used to construct a non-Gaussian density in laser gyroscope applications \cite{vogel_skewed_1987}.
More generally, where an exact expression for $\hat{\rho}$ is
unavailable, an approximate solution can often be estimated; for example, in
exponential and quadratic integrate-and-fire systems
\cite{richardson_firing-rate_2007}. This approximate solution can be used to
obtain an approximate mean first-passage time. Therefore, the approach to cast
statistics of nonlinear stochastic oscillators in a form of a Fredholm equation
allows analysis of the effects of correlated environmental noise in
a diverse range of problems.

\vspace{1mm}

\begin{acknowledgments} FF's work was supported by the Deutsche
	Forschungsgemeinschaft (Grant No. FA 1316/2-1).  CvW has received
	funding via CRCNS Grant No. ANR-14-NEUC-0001-01, ANR Grant No.
	ANR-13-BSV4-0014-02, and No. ANR-09-SYSC-002-01.  \end{acknowledgments}

\bibliography{ref}

\end{document}